\documentstyle[twoside]{article}
\DeclareSymbolFont{rsfs}{U}{rsfs}{m}{n}
\DeclareSymbolFontAlphabet{\mathscr}{rsfs}
\def\dfrac#1#2{{\displaystyle{\frac{#1}{#2}}}}
\catcode`\@=11
\newif\ifinany@
\def\onlydmatherr@#1{\Err@{\string#1\space allowed only in display math mode}}
\def\DN@{\def\next@}
\def\split{\relax\ifinany@\let\next@\insplit@\else
 \ifmmode\ifinner\def\next@{\onlydmatherr@\split}\else
 \let\next@\outsplit@\fi\else
 \def\next@{\onlydmatherr@\split}\fi\fi\next@}
\def\insplit@{\global\setbox\z@\vbox\bgroup\vspace@\let@\ialign\bgroup
 \hfil\strut@$\m@th\siplaystyle{##}$&$\m@th\displaystyle{{}##}\hfill\crcr}
\def\outsplit@#1$${\align\insplit@#1\endalign$$}
\def\theequation{\thesectionc.\arabic{equation}}

\catcode`\@=11
\long\def\@makefntext#1{
\protect\noindent \hbox to 3.2pt {\hskip-.9pt  
$^{{\eightrm\@thefnmark}}$\hfil}#1\hfill}		

\def\thefootnote{\fnsymbol{footnote}}
\def\@makefnmark{\hbox to 0pt{$^{\@thefnmark}$\hss}}	
	
\def\ps@myheadings{\let\@mkboth\@gobbletwo
\def\@oddhead{\hbox{}
\rightmark\hfil\eightrm\thepage}   
\def\@oddfoot{}\def\@evenhead{\eightrm\thepage\hfil
\leftmark\hbox{}}\def\@evenfoot{}
\def\sectionmark##1{}\def\subsectionmark##1{}}

\oddsidemargin=\evensidemargin
\renewcommand{\thefootnote}{\fnsymbol{footnote}}

\newcounter{sectionc}\newcounter{subsectionc}\newcounter{subsubsectionc}
\renewcommand{\section}[1] {\vspace{12pt}\addtocounter{sectionc}{1} 
\setcounter{subsectionc}{0}\setcounter{subsubsectionc}{0}
\setcounter{equation}{0}\noindent 
	{\tenbf\thesectionc. #1}\par\vspace{5pt}}
\renewcommand{\subsection}[1] {\vspace{12pt}\addtocounter{subsectionc}{1} 
	\setcounter{subsubsectionc}{0}\noindent 
	{\bf\thesectionc.\thesubsectionc. {\kern1pt \bfit #1}}\par\vspace{5pt}}
\renewcommand{\subsubsection}[1] {\vspace{12pt}\addtocounter{subsubsectionc}{1}
	\noindent{\tenrm\thesectionc.\thesubsectionc.\thesubsubsectionc.
	{\kern1pt \tenit #1}}\par\vspace{5pt}}

\newcounter{appendixc}
\newcounter{subappendixc}[appendixc]
\newcounter{subsubappendixc}[subappendixc]
\renewcommand{\thesubappendixc}{\Alph{appendixc}.\arabic{subappendixc}}
\renewcommand{\thesubsubappendixc}
	{\Alph{appendixc}.\arabic{subappendixc}.\arabic{subsubappendixc}}

\renewcommand{\appendix}[1] {\vspace{12pt}
        \refstepcounter{appendixc}
        \setcounter{figure}{0}
        \setcounter{table}{0}
        \setcounter{lemma}{0}
        \setcounter{theorem}{0}
        \setcounter{corollary}{0}
        \setcounter{definition}{0}
        \setcounter{equation}{0}
        \renewcommand{\thefigure}{\Alph{appendixc}.\arabic{figure}}
        \renewcommand{\thetable}{\Alph{appendixc}.\arabic{table}}
        \renewcommand{\theappendixc}{\Alph{appendixc}}
        \renewcommand{\thelemma}{\Alph{appendixc}.\arabic{lemma}}
        \renewcommand{\thetheorem}{\Alph{appendixc}.\arabic{theorem}}
        \renewcommand{\thedefinition}{\Alph{appendixc}.\arabic{definition}}
        \renewcommand{\thecorollary}{\Alph{appendixc}.\arabic{corollary}}
        \renewcommand{\theequation}{\Alph{appendixc}.\arabic{equation}}
        \noindent{\tenbf Appendix \theappendixc #1}\par\vspace{5pt}}
\newcommand{\subappendix}[1] {\vspace{12pt}
        \refstepcounter{subappendixc}
        \noindent{\bf Appendix \thesubappendixc. {\kern1pt \bfit #1}}
	\par\vspace{5pt}}
\newcommand{\subsubappendix}[1] {\vspace{12pt}
        \refstepcounter{subsubappendixc}
        \noindent{\rm Appendix \thesubsubappendixc. {\kern1pt \tenit #1}}
	\par\vspace{5pt}}

\newcommand{\textlineskip}{\baselineskip=13pt}
\newcommand{\smalllineskip}{\baselineskip=10pt}

\def\eightcirc{
\begin{picture}(0,0)
\put(4.4,1.8){\circle{6.5}}
\end{picture}}
\def\eightcopyright{\eightcirc\kern2.7pt\hbox{\eightrm c}} 

\newcommand{\copyrightheading}[1]
	{\vspace*{-2.5cm}\smalllineskip{\flushleft
	{\footnotesize submitted to the International Journal of Theoretical and Applied 
	Finance #1}\\
	{\footnotesize
	World Scientific Publishing
	 Company and Imperial College Press}\\
	 }}

\def\abstracts#1#2#3{{
	\centering{\begin{minipage}{4.5in}\baselineskip=10pt\footnotesize
	\centerline{\footnotesize ABSTRACT}
	\parindent=0pt #1\par 
	\parindent=15pt #2\par
	\parindent=15pt #3
	\end{minipage}}\par}} 

\def\keywords#1{{
	\centering{\begin{minipage}{4.5in}\baselineskip=10pt\footnotesize
	{\footnotesize\it Keywords}\/: #1
	 \end{minipage}}\par}}

\newcommand{\bibit}{\nineit}
\newcommand{\bibbf}{\ninebf}
\renewenvironment{thebibliography}[1]
	{\frenchspacing
	 \ninerm\baselineskip=11pt
	 \begin{list}{[\arabic{enumi}]}
	{\usecounter{enumi}\setlength{\parsep}{0pt}
	 \setlength{\leftmargin 13.7pt}{\rightmargin 0pt} 
	 \setlength{\itemsep}{0pt} \settowidth
	{\labelwidth}{[#1]}\sloppy}}{\end{list}}

\newcounter{itemlistc}
\newcounter{romanlistc}
\newcounter{alphlistc}
\newcounter{arabiclistc}
\newenvironment{itemlist}
    	{\setcounter{itemlistc}{0}
	 \begin{list}{$\bullet$}
	{\usecounter{itemlistc}
	 \setlength{\parsep}{0pt}
	 \setlength{\itemsep}{0pt}}}{\end{list}}

\newcommand{\fcaption}[1]{
        \refstepcounter{figure}
        \setbox\@tempboxa = \hbox{\footnotesize Fig.~\thefigure. #1}
        \ifdim \wd\@tempboxa > 5in
           {\begin{center}
        \parbox{5in}{\footnotesize\smalllineskip Fig.~\thefigure. #1}
            \end{center}}
        \else
             {\begin{center}
             {\footnotesize Fig.~\thefigure. #1}
              \end{center}}
        \fi}

\newcommand{\tcaption}[1]{
        \refstepcounter{table}
        \setbox\@tempboxa = \hbox{\footnotesize Table~\thetable. #1}
        \ifdim \wd\@tempboxa > 5in
           {\begin{center}
        \parbox{5in}{\footnotesize\smalllineskip Table~\thetable. #1}
            \end{center}}
        \else
             {\begin{center}
             {\footnotesize Table~\thetable. #1}
              \end{center}}
        \fi}

\def\pmb#1{\setbox0=\hbox{#1}
	\kern-.025em\copy0\kern-\wd0
	\kern.05em\copy0\kern-\wd0
	\kern-.025em\raise.0433em\box0}

\def\fnt#1#2{\footnotetext{\kern-.3em
	{$^{\mbox{\scriptsize #1}}$}{#2}}}

\def\fpage#1{\begingroup
\voffset=.3in
\thispagestyle{empty}\begin{table}[b]\centerline{\footnotesize #1}
	\end{table}\endgroup}

\font\tenrm=cmr10
\font\tenit=cmti10 
\font\tenbf=cmbx10
\font\bfit=cmbxti10 at 10pt
\font\ninerm=cmr9
\font\nineit=cmti9
\font\ninebf=cmbx9
\font\eightrm=cmr8

\newtheorem{thm}{\indent Theorem}

\newtheorem{lem}{Lemma}

\newtheorem{prop}{Proposition}
\newtheorem{obs}{}

\newcommand{\proof}[1]{{\tenbf Proof.} #1 $\Box$.}

\def\qed{\hbox{${\vcenter{\vbox{			
   \hrule height 0.4pt\hbox{\vrule width 0.4pt height 6pt
   \kern5pt\vrule width 0.4pt}\hrule height 0.4pt}}}$}}

\renewcommand{\thefootnote}{\fnsymbol{footnote}}	

\newcommand\bit{\begin{itemlist}}
\newcommand\eit{\end{itemlist}}
\def\comma{\,\,\,\,\,,\,\,\,\,\,}
\def\bea{\begin{eqnarray}}
\def\eea{\end{eqnarray}\noindent}
\def\al{\alpha}

\def\si{\sigma}

\def\Ga{\Gamma}
\def\Th{\Theta}
\def\La{\Lambda}

\def\De{\Delta}

\begin{document}

\normalsize\textlineskip
\thispagestyle{empty}
\setcounter{page}{1}

\copyrightheading{}			

\vspace*{0.88truein}

\fpage{1}
\centerline{\bf ON THE CONSISTENCY OF THE DETERMINISTIC}
\vspace*{0.035truein}
\centerline{\bf LOCAL VOLATILITY FUNCTION MODEL}
\vspace*{0.035truein}
\centerline{\bf (`IMPLIED TREE')}
\vspace*{0.37truein}
\centerline{\footnotesize KARL STROBL\footnote{e-mail: karl.strobl@bigfoot.com}}
\vspace*{0.015truein}
\centerline{\footnotesize\it ABN AMRO Bank N.V.}
\baselineskip=10pt
\centerline{\footnotesize\it 250 Bishopsgate, London EC2M 4AA, U.K.}
\vspace*{1truein}
\abstracts{We show that the frequent claim that the implied tree prices
exotics consistently with the market is untrue if the local volatilities
are subject to change and the market is arbitrage-free.
In the process, we analyse -- in the most general
context -- the impact of stochastic variables on the P\&L of a hedged
portfolio, and we conclude that no model can {\it a priori} be expected
to price all exotics in line with the vanilla options
market. Calibration of an assumed underlying
process from vanilla options alone must not be overly restrictive, yet still
unique, and relevant to all exotic options of interest.
For the implied tree we show that the calibration to real-world prices
allows us to only price vanilla options themselves correctly.
This is usually attributed to the incompleteness of the market
under traditional stochastic (local) volatility models. We show that
some `weakly' stochastic volatility models without quadratic variation of
the volatilities avoid the incompleteness problems,
but they introduce arbitrage.
More generally, we find that any stochastic tradable either has quadratic
variation -- and therefore a $\Ga$-like P\&L on instruments with
non-linear exposure to that asset -- or it introduces arbitrage
opportunities.}{}{}

\vspace*{5pt}
\keywords{Implied trees; Stochastic local volatility; Exotic options;
Vega hedging; Skew models; Market incompleteness;}


\textheight=7.8truein
\setcounter{footnote}{0}
\renewcommand{\thefootnote}{\arabic{footnote}}

\vspace*{1pt}\textlineskip	
\section{Introduction}	
\vspace*{-0.5pt}

One of the customary ways in dealing with the volatility smile and
term-structure as observed, for example, in the implied Black-Scholes
volatilities in the options markets, is to translate the different
Black-Scholes volatilities into a `local' volatility (`implied tree')
\cite{1,2,3}. This assumes the stock-price follows the process
\[
\frac{dS}{S}=\mu(t)dt+\sigma(S(t),t)dW,
\]
where $\sigma$ is a deterministic function of the stock-price and time,
calibrated such that the model matches the observed vanilla options prices,
and $W$ is a standard Brownian motion (SBM) and the only random source of
the model.
While the earliest incarnations of this idea
led to very erratic behaviour of the local volatility surface, much recent
work has focused on how to compute a surface that is both
reasonably smooth and fast to find \cite{4,5,6,7}.
Both the names DVF (deterministic
volatility function) and LVF (local volatility function) are customary for
this approach. We will use the abbreviation DLVF, for obvious reasons, and to 
distinguish it from SLVF (stochastic local volatility function, or stochastic
implied tree models), which
we shall also discuss in this paper. We shall use LVF when we refer to
both classes of models.

DLVF models are the preferred skew models of a number of
institutions, given that alternative approaches to create a skew effect
encounter limitations as to how well they can fit market data
\cite{8}. The commonly perceived downside of DLVF models are practical
limitations in measuring the volatility
exposure of an exotic options portfolio.
There are, however, no theoretical grounds on which to
affirm that volatility risk can be hedged at all in this
framework: A truly deterministic volatility function 
must be assumed. The two most pressing theoretical reservations one may
harbour towards local volatility models, and the problems arising from
these concerns, are therefore the following.
\bit
\item
When the stock-price moves, but the LVF is
assumed to be constant, this imposes a rule on the Black-Scholes-implied
volatilities which is neither of the two popular rules nick-named
`sticky strike' and `sticky delta'. The `sticky-implied-tree'
rule leads to different stock-deltas than either of the other two rules
\cite{9}. Therefore, before putting trust in local volatility models, we 
must either justify the emergence of this rule, or show that the validity
or otherwise of the `sticky-implied-tree' assumption is immaterial to
the P\&L of a hedged strategy, which can only be the case if we can hedge
against stock and volatility movements separately.
\item
In general, it is difficult to estimate how the local volatility
surface moves when a limited number of Black-Scholes-implied volatilities
change in the market. Note, however, that measuring this exposure is one of
the primary motivations for using a skew model at all.
One of the key arguments used in defence of DLVF models is that volatility
risk is already priced into the market, and that the DLVF model is only
an {\it effective theory} which makes use of the fact that volatility risk
is already integrated out in the right way by the market-implied local
DLVF. This is true for a well-diversified portfolio
of vanilla options, but the implied volatility does not compensate
sufficiently for {\em option-specific} vega risk of vanilla or exotic options.
A $\De$-hedged
portfolio short a log-contract has a much more predictable P\&L in the
real world than a $\De$-hedged portfolio short just one at-the-money vanilla
option. This is because in the first case the unknown P\&L component is
only a function of the realised asset price variance (assuming a diffusive
asset price process), whereas in the second case the P\&L is also very much
a function of the average $\Ga$ experienced over the life of the option.
So, although implied volatilities often compensate the option seller for the
vega-risk taken, they do so insufficiently for someone who does
not diversify his volatility risk across various strikes. If we want to
sell an exotic option, and hope that it is priced in line with the market,
we must also show that we either can diversify our vega exposure
in a similar way, and that the vega-risk which can not be diversified away
is priced in line with the vega-risk premium that is already integrated out
in the observable Black-Scholes implied volatilities. Obviously, the DLVF model
does not give us a volatility risk premium above the premium implied in
the vanilla prices. This is what should be meant when it is claimed
that the implied tree values exotics `in line with the
market'. The concern that needs to be addressed is thus whether the
expectations that are already integrated out in the `vanilla-implieds'
can be re-used as `exotic-implieds'.
Note that the only way we can be certain that the effective
integration of volatility risk can be applied to exotics is by showing
that volatility risk of an exotic option can be hedged with a dynamic
trading strategy of vanilla options. Unless one can 
show that volatility exposure can be hedged against any of the
points on the local volatility surface (or any of the variables which
parameterise this surface), we may not be entitled to use risk-neutral
valuation techniques in a world where implied volatilities are
subject to change. This would make the DLVF models unviable.
\eit
We will demonstrate in this article that these two concerns can be
redressed in an arbitrage-free market only if the future
spot-volatilities are indeed deterministically identical to their DLVF values.
That they are not is clear from empirical work and from simple
intuition \cite{10,11}.

The structure of the paper is as follows: First, we give a general
discussion of what conditions we might expect in order for the local
volatility model to be viable under risk-neutral valuation with an
only mildly time-varying volatility function.
We observe that these requirements destroy the no-arbitrage condition,
and that we thus need to consider the full SLVF model known also as the
stochastic implied tree. This paper can
be described as a practical guide to (a) the problem of incompleteness
and (b) the HJM-like no-arbitrage condition of stochastic implied trees
as described in \cite{12}. The above-mentioned early sections of the paper
introduce the theoretical problems. Although the way of presenting these
problems may be new, the incompleteness problem has been discussed in
much greater detail and mathematical rigor in the literature on
stochastic volatility models.

After this practitioner's guide to the theoretical problems, we 
present a discrete framework in
which we are able to construct, if it exists, the replicating trading
strategy of any exotic option, involving stock and Arrow-Debreu option
positions. We demonstrate that this portfolio always exists in a two-step
trinomial world. Next, we analyse a three-step trinomial world and find
that the stochastic process required to make the hedge work violates the
no-arbitrage condition. This problem persists in the continuum limit.

We then present the conclusion that in a DLVF model
risk-neutral valuation of exotics (without additional arbitrary parameters like
for example option-dependent volatility risk premia, which can not be
calibrated to fit the vanilla options market) is inconsistent with the
assumptions of no arbitrage and unknown future volatility.

\section{Stochastic Processes with and without Quadratic Variation}
\label{sec:quadratic_variation}

Some of the issues discussed here benefit from a brief tour through
familiar territory. The Black-Scholes world is based on the
assumption that the stock price $S$ follows the process
\[
dS=rSdt+\si S dW,
\]
where the $dW$ are the increments of an SBM. The Black-Scholes PDE
\[
\frac{\partial f}{\partial t}+rS\frac{\partial f}{\partial S}+
\frac{1}{2}\sigma^2 S^2 \frac{\partial^2 f}{\partial S^2}=rf,
\]
in a slightly different notation, is called the {\em continuous trading
equation}
\[
\Theta \, dt+rS\De\, dt+\frac{1}{2}\sigma^2 S^2 \Gamma dt=fr\,dt.
\]
The left-hand side of
the continuous trading equation is equal to $df-\De dS+r\De Sdt$ , the
P\&L of a $\De$-hedged portfolio,
which is risk-less in the Black-Scholes world, because the volatility
$\sigma$ is a known constant. For the rest of the discussion, it will be
easier to think of the continuous trading equation in terms of a
futures-hedge, as the funding term then drops out, i.e.~we think of the
continuous trading equation as
\[
df-\frac{\partial f}{\partial F}dF=rfdt,
\]
or, after expanding the total differential for $df(F,t)$ to first order in
$dt$,
\[
\Th dt+\frac{1}{2}\sigma^2 F^2 \Ga_F dt=rfdt.
\]
The continuous trading equation thus tells us that if all the other
assumptions of the Black-Scholes world hold, the instantaneous P\&L of a
long, $\De$-hedged, option position is a trade-off between the loss of
$\Th$ on the option and the gain on $\Ga$, which is proportional to the
instantaneous variance rate $\sigma^2$ of the stock-price. The fact
that we have a $\Ga$-related P\&L component is a consequence of the fact
that the stock-price has non-vanishing quadratic variation $dS,dF={\mathscr{O}}
(\sqrt{dt})$.
Now imagine that $S$ is stochastic, but of vanishing quadratic variation,
for example $dS\equiv\left(r+W-\frac{t^2}{2}\right)Sdt$ or
$dF\equiv\left(W-\frac{t^2}{2}\right)Fdt$, then the
left-hand side of the continuous trading equation is $df-\De_F dF$ or
\[
\Th dt=rfdt,
\]
so the $\Ga$-related P\&L disappears, even though the asset price process is
stochastic. The asset price process has a fundamental problem, however:
Even though initially it has the desirable property of
$\hat{E}\left[F_t|F_0\right]=F_0$ \footnote{~To prove this, simply use the
fact that the random variable $X_t=\int_0^tW(u)du$ with $W(0)=0$ is normally
distributed with zero mean and variance $\frac{t^3}{3}$.}, it is in general
not the case that $\hat{E}\left[F_t|F_u\right]=F_u\,\,,\,\,u>0$, so
the futures price is not a martingale:
because $\frac{\partial S}{\partial t}$ persists for some time, observing
the `momentum' of the stock price would provide arbitrage opportunities
between stock-returns and funding costs.
In this model, options are priced at their
discounted intrinsic value, which is consistent with the observation that
holding $\Ga$ is of no value. In reality, people pay a price for the privilege
of being long $\Ga$ and the $\Th$ is rather smaller than the cost of
funding on the option premium and often negative, which thus indicates that
the market believes that
the stock price does have higher than linear variations.
This means that the option market `knows' that the stock price is
either diffusive or exhibits jumps, or both. In
turn, this also means that the arbitrage opportunity related to the
persistence of the stock-momentum disappears. To state this formally:
\begin{prop}
\label{prop:1}
For the future $F$ of an asset $S$ (in a world of
deterministic funding costs) which follows the Ito process
\[
dF=\left(\xi B(t)-\frac{\xi^2 t^2}{2}\right)F\,dt+\sigma(F,t)F\,dW
\] 
where $B(t)$ and $W(t)$ are SBMs, $F,\xi,\sigma\ge 0$,
and the entire information on the asset price is it's history, the following
statements are equivalent
\begin{enumerate}
\item $B(t)$ is observable
\item $\sigma(F,t)\equiv 0$ and $\xi\ne 0$.
\item $\xi\ne 0$, and the market satisfies, at any given time
$B(t)\stackrel{>}{<}\frac{\xi t^2}{2}\Longleftrightarrow  dF\stackrel{>}{<} 0$.
\item The future price is predictable to order $dt$, and, except when
$B(t)=\frac{\xi t^2}{2}$, allows arbitrage.
\item The price process $F(t)$ is stochastic and differentiable almost
everywhere. It thus has no quadratic or higher variation.
\end{enumerate}
Also, if any of the conditions is true, there can be no risk-neutral
measure for $F$.
\end{prop}
\proof{
$1\Longleftrightarrow2$ is trivial, as no two Gaussian variables can be
simultaneously inferred from one futures price. $2\Longleftrightarrow3$ follows
from the fact that, if $\sigma\ne 0$, almost all realisations of
$dW$ would dominate over the $dt$-term in $dF$, so the
statements in item 3 could not be made with any certainty, and vice versa.
The first part of $2\Longleftrightarrow4$ is trivial, since the leading order
in $dF$ is $dt$ with coefficients which are (by equivalence with item 1)
observable, if conditions $2$ are satisfied. This, however, would still
be true if $\xi$ were zero. However, in this 
case $dF\equiv 0$, and the market would not allow arbitrage. Yet if $\xi\ne 0$,
then we know the sign of $dF$, and arbitrage exists, in that a zero-price
position in $F$ can be acquired that is instantaneously profitable. Thus also
$4\Longrightarrow2$. $5\Longrightarrow2$ can be shown in the following
way: if $\sigma\ne 0$ anywhere,
the price process would have quadratic variation, as there are terms of order
$\sqrt{dt}$. If $\sigma\equiv 0$ and $\xi\equiv 0$, however, the process is
deterministic. Thus a stochastic price process without quadratic variation
must have $\si\equiv 0$ and $\xi\ne 0$. $2\Longrightarrow5$ is then
just the trivial reversal of this argument.

The non-existence of the risk-neutral measure follows directly from the
existence of arbitrage opportunities, according to Harrison and Pliska
\cite{13}}

Note also that the observation, originally made in the HJM framework
for interest rates, that the no-arbitrage condition determines
the drift of a stochastic, tradable variable is related to this proposition.
This observations was extended to stochastic local volatilities by Derman
and Kani \cite{12,14}.
For both interest rates and stochastic volatilities it is also true that
in the absence of quadratic or higher variations (i.e.~zero volatility in
a jump-less diffusion model) the no-arbitrage drift is zero. Proposition 1
only applies to a specific type of process, but makes no assumptions
as to the nature of the tradable asset.

Let us therefore assume the somewhat more general jump-less, stochastic
futures price process
\begin{equation}
dS(t)=\left(r(t)+I(t)-
\frac{\partial}{\partial t}\ln\hat E_0\left[e^{\int_0^tI(u)du}
\right]\right)S(t)dt+\si (W,s)S(t)dW(s),
\label{eq:I}
\end{equation}
where $I(t)$ is an Ito process, and $\hat E_0$ is the expectation operator
evaluated at time zero. Henceforth, we call a process of the type
Eq.~(\ref{eq:I})
a {\it generalised Ito process}, which differs from the usual
Ito process because the drift is an Ito process itself, but with a
deterministic drift correction
$-\frac{\partial}{\partial t}\ln\hat E_0\left[e^{\int_0^tI(u)du} \right]$.
Funding rates are deterministic, such that the futures price $F(t)$
follows the same process, but without the interest
rate drift component. This then has the solution
\[
F(t)=F_0\dfrac{\exp\left[\int_0^tI(s)ds+\int_0^t\si(F(s),s)dW-
\frac{1}{2}\int_0^t
\si^2(F(s),s)ds\right]}{\hat E\left[\exp\int_0^tI(s)ds\right]},
\]
which has similar properties to the process of Proposition \ref{prop:1},
but is more general.
This leads to a natural generalisation of Proposition \ref{prop:1}:
\begin{thm}
\label{thm:1}
For a future -- in a world of deterministic funding costs -- on an asset,
which follows the process Eq.~(\ref{eq:I}),
the following statements are equivalent
\begin{enumerate}
\item $\si(F,t)\equiv 0$ and $I(t)$ is stochastic.
\item $I(t)$ is stochastic, and, at any given time
$I(t)\stackrel{>}{<}\frac{\partial}{\partial t}\ln\hat E_0\left[\exp
\int_0^t I(s)ds\right]\Longleftrightarrow dF\stackrel{>}{<} 0$
\item The future price is predictable to order $dt$, and, except when
$\frac{\partial F}{\partial t_{-}}=0$, allows arbitrage.
\item The price process $F(t)$ is stochastic without quadratic variation.
\end{enumerate}
Furthermore, if these conditions are true, there is no risk-neutral
measure for $F$.
\end{thm}
\proof{
As opposed to Proposition \ref{prop:1}, here
we have to leave out the observability of the realised value of
$I(t)$, as this would require knowledge of
$\frac{\partial}{\partial t}\ln\hat E_0\left[\exp\int^t I(s)ds\right]$,
which can not be taken for granted. $1\Longleftrightarrow 2$ is analogous to 
Proposition \ref{prop:1}, and items 3 and 4 follow immediately}

The important lesson to learn from Theorem \ref{thm:1}
is that we do not need to
know the details of the generalised Ito process that makes $F$
stochastic in order
to conclude that in the absence of quadratic variation there are arbitrage
opportunities which we can spot simply by observing
$\frac{\partial F}{\partial t}$, and the only way to exclude arbitrage at
all time is to make $F(t)$ deterministic and $\hat E_t\left[dF_u\right]=0\,\,
\,\,\forall u>t$,
which of course means that $F(t)$ is simply a constant in time.

From this we introduce the important
\begin{lem}
If, in a world of deterministic funding costs, a tradable
asset follows a process with no quadratic or higher variation, it is either
deterministic or allows arbitrage almost always, or both.
\label{lem:1}
\end{lem}
\proof{
The fact that the process has no higher than quadratic variations and that
the futures price is a martingale means
that it is a generalised Ito process of the type Eq.~(\ref{eq:I}),
except that the drift term $I(t)$ in Theorem \ref{thm:1} can be generalised
to a jump-diffusion process where the set of times where jumps occur are 
of measure zero (i.e.~the jumps are always countable, like in a Poisson
process). Then the quadratic variation of the process $F(t)$ is zero, and
the price-process is almost always predictable at least to order $dt$.
This means it is differentiable almost everywhere, and continuous
everywhere. The differentiability almost everywhere would allow
successful arbitrage almost always (with infinitesimal losses in
the countable cases where it is not differentiable), if the slope of
the process is different from the funding costs on the asset (because the
cost of setting up a futures position is zero).
Therefore, the process allows no arbitrage only if $F$ is constant almost
everywhere.
Furthermore, if the generalised Ito process of no quadratic variation does
allow arbitrage, it is very easy to spot the arbitrage by observing the
momentum of the futures price}

It is very important to note that we never implied in any of the proofs of
the above theorem and lemma that the future is written on a stock-price:
It could have been a local volatility, as long as the funding costs are
deterministic. Again, this is a generalisation of the observations of
HJM, and Derman and Kani to any stochastic tradable
asset under zero volatility \cite{12,14}.
The HJM drifts for interest rates and the local-volatility drifts as
calculated by Derman and Kani prove that even for non-zero volatility the
drifts must be deterministic, if the no-arbitrage condition is to be
satisfied. Lemma \ref{lem:1} is thus also more restricted
than the results of \cite{12,14}, in that we only analyse the zero-volatility
case, but it is also more general, in that it applies to any tradable asset.

The bit to keep in mind for the next sections is that a stochastic tradable
has to have at least quadratic variation in order to preclude arbitrage.

\subsection{Does the quadratic variation matter in discrete hedging?}
\label{sec:2.a}

Against taking quadratic variation too seriously, one can frequently hear
variations of the following objection:

`Quadratic variation has to do with the fractal
dimension of the process. It basically says, if I add the moduli of all
the stock-price changes in any finite interval, I get infinity. This is
patently not the case in reality. Furthermore, I typically re-hedge only
after a stock-move of, say, 4\%. So I re-hedge maybe once a week on average.
Adding all the moduli of stock-returns relevant to my hedging-strategy, 
they obviously don't diverge. So for practical purposes, the stock-movement
relevant to me is proportional to time.
In other words: to me the stock-price path has dimensionality 1'.

It is easy to see what is wrong with this argument.
Our hedger experiences a typical $\Ga$-related P\&L
component of $52\cdot(0.04)^2\Ga$ on an annualised basis. The crux is that,
if his pain-barrier was a stock-move of, say, 2\%, he would not re-hedge
once a week, or even twice a week on average, but four times a week,
if the quadratic variation 
of the stock-price process is the highest non-vanishing variation.
So on an annualised basis, his $\Ga$-related P\&L component is
$208\cdot(0.02)^2\Ga$, which is exactly the same as before. The quadratic
variation therefore does matter equally for someone who hedges infrequently
as for someone who hedges continuously.  One can not escape the $\Ga$-related
P\&L by simply re-hedging more frequently. In other words: the fractal
dimension of the stock-price process matters regardless of the granularity
of price observations. If our hedger lived in a world where the
dimensionality of the stock-price process is linear, dropping his
pain-barrier to 2\% would incur a $\Ga$-related P\&L component of
$104\cdot(0.04)^2\Ga$, and he could observe, regardless of the granularity
of his hedge, that more frequent re-hedging lets him reduce the $\Ga$-P\&L.
Let us therefore not be fooled: The fractal dimension of the stochastic
process is observable. In particular: \begin{thm}The fractal dimension on the
scale relevant to discrete-time hedging is also observable.\label{thm:3}
\end{thm}

\subsection{Hedging `Exotics' in the Fixed-Income World}

Before we tackle the stochastic local volatility world, let us consider an
instructive example of the typical problem we discussed in section 2 from the
fixed-income markets.
Kani {\it et.~al.}~observed that trading and hedging against forward
interest rates is similar to trading and hedging against local volatilities
\cite{15}.
This allows us to use examples which are easy to follow
because the set of independent forward rates is one-dimensional, while the
set of local volatilities is two-dimensional. Here we bring an example
of a very simple fixed income world with only two bond-prices, and a
simple `exotic' (i.e.~a simple function of these two prices) which can not
be replicated with a dynamic trading strategy without taking new
independent variables into account, namely variance and covariance rates
of the interest rates.

This is a further qualificatnion of 
the claim by Kani {\it et.~al.}, namely
that everything in the fixed-income world can be replicated by dynamically
attainable `forward rate gadgets' which have zero cost and provide a specified
degree of exposure to a particular forward rate \cite{15}.
The central conclusion
we work towards here is the inapplicability of an effective DLVF theory to
an SLVF world. Consider thus the frequent claim that vanilla options make
local-volatility gadgets attainable, which complete the market even in the
presence of exotic options. This claim and the above-mentioned claim of
Kani {\it et.~al.}~are are easily mis-interpreted for the same reason,
namely that quadratic
variations are not negligible in a stochastic, arbitrage-free market, and
that they can not be fully hedged against \footnote{We would like to stress
that ref.~\cite{12} indicates that Kani {\it et.~al.}~are not vicitims of
this mis-interpretation. What matters here is only that ref.~\cite{15} could
easily mislead the reader into believing that forward-rate gadgets are perfect
hedging instruments, as the paper does not refer to the necessity of making
convexity corrections. The point we make is that misreading the paper in this
way is analogous to ignoring the quadratic variation of local volatility in
DLVF models.}.
Note, however, that this
observation can already be made in the derivation of the Black-Scholes
model: Perfect replication of a vanilla option in the Black-Scholes world
is only possible when we also know the volatility of the stock. For
multi-factor models, we need to know the variance-covariance rate matrix
in order to achieve perfect replication. The DLVF models make no reference
to this matrix, and there are usually not enough degrees of freedom to
calibrate to such a matrix if it were known.

Consider a world with only two
bonds $B_1$ and $B_2$, maturing at times $t_1$ and $t_2$ from now. The
corresponding interest rates are defined through
$B_1=\exp(-r_1t_1)$ and $B_2=\exp(-r_2t_2)$.
The forward rate is the interest 
rate for the term between $t_1$ and $t_2$ which can be locked in at
present, i.e.~$f=\frac{r_2t_2-r_1t_1}{t_2-t_1}$, therefore
$\frac{B_1}{B_2}=\exp\left(f\cdot(t_2-t_1)\right)$. The `interest rate gadget'
is the portfolio $\Lambda=B_1-\left[\frac{B_1}{B_2}\right]B_2$, where the 
square brackets indicate a position size (`hedge ratio'), which means a value
that stays fixed
during instantaneous changes of the market observables $r_1$ and $r_2$.
Now imagine we have some other instrument, whose exposure to the 
forward rate $f$ we want to hedge. We observe that
\[
\delta\Lambda=B_1\left[(t_2-t_1)\delta f-t_1(t_2-t_1)\delta f\delta r_1-
\frac{1}{2}(t_2-t_1)^2(\delta f)^2\right]\,.
\]
Kani {\it et.~al.}~mention that the bond $B_1$ can be synthesised
by holding one gadget and $\frac{B_1}{B_2}$ units of $B_2$, which gives exact 
replication to all orders \cite{15}.
Unfortunately, this is not a valid example of
hedging, because all we have in the
`replicating' portfolio is exactly one bond $B_1$. To really keep the
correspondence with DLVF models, we need to introduce a new instrument
which has a pay-off as a function of $f$, but which is not already in this
market. Consider for
example a security which pays exactly the forward rate as, say, a US dollar
amount. To hedge this, we need a strategy whose P\&L is and stays
linear in the forward rate. However, since the hedge ratio itself is a 
function of the forward rate, we get the second derivatives in
$\delta\Lambda$. If we hedge the forward-rate exposure of some portfolio
with the gadget $\Lambda$, we see that we introduce a negative P\&L component 
proportional to the variance rate of $f$, and a correlation term between
the forward rate and $r_1$. We do not have enough degrees of freedom to
eliminate these new terms, i.e.~to synthesise a product which is strictly
linear in the forward rate. The corresponding effect on, say, for example,
interest rate futures prices is frequently referred to in the fixed income
markets as a `convexity correction'. The size of the convexity correction
is increasing with the variance rate ${(df)^2}/{dt}$ and the covariance
rate ${(df)(dr_1)}/{dt}$. Note that if we know both these rates, perfect
replication is still possible, in the same sense as it is possible in the
Black-Scholes world for vanilla options when we know the volatility of
the underlying.

\section{Stochastic Local Volatilities}

Define $\vec x=(F_S,\vec\alpha)$, where $\vec\alpha$ is
an $n$-component vector parameterising the local
volatility surface, calibrated to market prices, and $F_S$ is the
forward price of the stock. Let us now consider a scalar function
$f(\vec{x})$ of these $n+1$ stochastic variables, and define the
vector-operator
\[
\vec\nabla\equiv\left(
\frac{\partial}{\partial F_S},
\frac{\partial}{\partial\alpha_1},
\frac{\partial}{\partial\alpha_2},\ldots,
\frac{\partial}{\partial\alpha_n}\right).
\]
Then
\[
df=f_tdt+(\vec\nabla f)\cdot(d\vec x)+\frac{1}{2}
(d\vec x)^T\cdot(\vec\nabla\otimes\vec\nabla f)\cdot (d\vec x) +\ldots
\]
If the stochastic process for $\vec{x}$ has no more than quadratic variation,
the higher order terms are smaller than order $dt$. Let us assume that
local volatility futures (or local volatility `gadgets' in the language
of ref.~\cite{15})
are attainable at zero cost. Then, after hedging the
portfolio to first order in the stock price and local volatilities, the
continuous trading equation is
\begin{equation}
f_tdt+ \frac{1}{2} (d\vec x)^T \cdot\left(\vec\nabla\otimes\vec\nabla f\right)
\cdot(d\vec x)=rfdt\,,
\label{eq:SLVFIIPL}
\end{equation}
which makes for a total of one $\Th$ component, the funding on the
stock-hedge, and $\frac{(n+1)n}{2}$ terms arising from all the different
second derivatives with respect to the different components of $\vec{x}$
\footnote{~The assumption that local volatility futures are attainable
is probably no more permissible than the assumption in the fixed income
world that forward rate gadgets are attainable, that is they are
attainable only to first order. In this sense, Eq.~(\ref{eq:SLVFIIPL})
will not be strictly correct, but will contain other second order
terms, arising from the second-order terms of the local volatility
gadgets. Qualitatively, the number of
P\&L terms is predicted correctly, however, as we only need replace
$\left(\nabla \otimes \nabla f\right)_{ij}$ with
$\left(\nabla \otimes \nabla f\right)_{ij}+C_{ij}$,
where $C_{ij}$ represents the combined second order exposure created
by our holding the local volatility gadgets.}.
We now have to make assumptions on what the asset price process is in the
real world. Broadly, we can classify the imaginable price processes
into four groups:
\begin{itemlist}
\item{DLVF world: }
The local volatilities are completely deterministic. The Black-Scholes
world is a trivial example. The asset price process of an implied tree is
the most general case.
\item{WSLVF world: }
The local volatilities are stochastic, continuous everywhere, and differentiable
almost everywhere. Let us call this
model the `weakly stochastic local volatility function' model.
\item{SLVF world: }
The variables parameterising the volatility function follow stochastic
processes with non-vanishing quadratic variation, but vanishing higher
variations (This means these processes are diffusive). Call this the
stochastic local volatility function model.
\item{Jump-diffusion world: }
Either the stock-price process or the volatility parameters (as implied by
the vanilla options market) or both are not simply diffusive. This would
manifest itself in higher than quadratic variations.
\end{itemlist}

Note that the decision on which world a particular market corresponds to
can be based on empirical data according to Theorem \ref{thm:3}.

Let us consider the suitability of the DLVF model to all of these four worlds.

In the DLVF world the local volatilities are perfect forecasts of
future realised instantaneous volatilities. In the absence of
arbitrage, the evolution of the local
volatility surface is static except for a shift of the time-frame of the
observer. In this case, risk-neutral valuation and the DLVF model
are obviously applicable. However, the viewpoint that we live in such a DLVF
world contradicts empirical evidence \cite{10,11}.

The WSLVF world has
the previously described arbitrage opportunities: observe the
local volatility `momentum', and simply go long the gadgets corresponding
to growing local vols, and short the others. This provides a risk-less
profit almost always \footnote{~A necessary condition for the existence of
such arbitrage is that the `almost never' occurring cases of making a loss
produce a bounded loss, which is the case here.}.
Within this world the implied tree does not work in the way it is
customarily used and clearly all vanilla options and all options calculated
on the implied tree are priced assuming a sub-optimal hedging strategy.
In fact there is no optimal hedging strategy: any given strategy
can be beaten by another one using more of the arbitrage
opportunities. Clearly, while the world could be WSLVF,
we gain nothing by making this assumption, unless we have a model
to identify and exploit the arbitrage opportunities.
We can make the intriguing observation, however, that everyone gets away with
using risk-neutral valuation as long as no market participant spots the
arbitrage. If option buyers and sellers all worked with a DLVF model and the
associated hedging strategies, the existence of arbitrage is irrelevant for
practical purposes. So it is in fact entirely possible that the real
world is of the WSLVF type, although, we would clearly prefer to have a
stronger cause for the use of risk-neutral valuation in the DLVF model than
the reason that it works simply because everybody is using it. Note that the
assumption that the real world is WSLVF is easily testable: use a DLVF model,
and take local volatility positions according to the `momentum' of the
local volatility parameters. Unless one makes profits almost always with
this strategy, the real world is not WSLVF.

Next we have to discuss the SLVF world. A necessary condition for local
volatility gadgets to exist is that tradability of all
European vanilla options, which is equivalent with the tradability of all
Arrow-Debreu options. Since the existence of sufficiently many of these
options is a necessary condition to calibrate the LVF surface, we will assume
in the remaining part of this paper that sufficiently many Arrow-Debreu
options or European vanilla options are available in order to produce
gadgets for all the LVF parameters we allow in the model, and that these
options can be used in any hedging strategy for exotic options.
Then we have two possibilities to distinguish:
\bit
\item{SLVF(I) world:}
Local volatility gadgets exist, but second-order exposure can not be
hedged. The gadgets themselves may or may not produce additional
second-order exposure.
\item{SLVF(II) world:}
In this world, we can hedge against
variations of stock price and local volatilities up to second order.
\eit
The central issue of this paper is that the SLVF(II) world is theoretically
impossible unless there is a very low-parametric representation of the LVF,
and that, in any case, volatility of volatility has P\&L effects which accrue
over time.

We do not consider the jump-diffusion world for two reasons. Firstly, we are
not aware of any specific claim that DLVF models could be appropriate to such
a world, although the idea that effective theories may also have jump-risk
`effectively' integrated out can not be discarded {\it a priori}.
Secondly, however,
we already show that there irreconcilable problems with the application of
DLVF models in an SLVF world. Since jump-diffusion models are a super-set
of SLVF models, we would have to hedge against even more sources of
uncertainty in jump-diffusion worlds. It is thus unnecessary to prove that
the problems of the SLVF worlds persist in the jump-diffusion setting.

\section{P\&L in the SLVF worlds}
\label{sec:SLVFworlds}

In an SLVF(I) world, we have all the P\&L terms
of Eq.~(\ref{eq:SLVFIIPL}) to consider. This will lead to risk premia
on the option proportional to the variance of the to-be-realised
`$\Ga$' P\&L, in other words proportional to
\begin{equation}
d \vec x^T\cdot\int_0^T\left(\nabla\otimes\nabla f\right)\cdot d\vec x,
\label{eq:secondorderterms}
\end{equation}
or simply to all the un-hedged second order terms, because
they are of order $dt$ and thus accrue P\&L proportional with time.

The question is now whether we can quantify the volatility risk
premium appropriate for an exotic option in an SLVF(I) model.
Obviously, the terms in Eq.~(\ref{eq:secondorderterms}) are in principle
computable, if the market-implied covariances between all local
volatilities are known, as the rest is simply
path-integration \footnote{~We make no statement about the practical
difficulties of attempting this
computation.}. There is of course no hope to infer these parameters
unless the {\it a priori} parameterisation of the LVF we are willing to
consider is extremely low-dimensional: For an $n$-dimensional
parameterisation, we have to infer the $n$ parameters, and all components
of the variance-covariance matrix of these parameters plus the stock-price.
This is a symmetric $(n+1)\times(n+1)$ matrix. In total, we have thus
$\frac{(n+1)(n+2)}{2}+n$
parameters to estimate (in a non-mean-reverting SLVF model).
Assuming we have $m$ independent vanilla
option prices, we are allowed to use no more than
$\frac{\sqrt{41+8m}-7}{2}$ parameters for the LVF. We thus need four
vanilla option prices for the usual non-mean-reverting stochastic volatility
model, and at least eight option prices to accommodate even the simplest
skew information, by specifying the SLVF by two volatility parameters.
If the SLVF
is required to be mean reverting, we require twelve options to calibrate
such a model. Now say one of those parameters was designed to capture 
skew, and the other to capture term structure.
If we therefore took a second skew parameter, making the total parameterisation
of the LVF three-dimensional, we would require 19 options in the
mean-reverting case and 13 otherwise. This analysis underlines the
urgent need for `smoothing' or `regularising' of the LVF, which has been
attempted by many authors recently \cite{4,5,6,7}.
However, smoothing alone does not suffice to estimate volatility risk premia.

At this point in the discussion, one might want to ask the question of
whether the volatility risk premium really matters. After all, we never
worry about it in the Black-Scholes world, even though the Black-Scholes
assumption of deterministic volatility is clearly wrong in practice.

\subsection{Volatility Risk Premia and Exotic Options}
\label{sec:4.1}

It is in the context of Black-Scholes type models, that we
find that there is skew: The BS volatilities
implied in the market differ for different strikes and maturities.
This is ascribed to the BS assumptions being wrong in one way or another.
The LVF approach is to say that volatility simply is not flat, the
stochastic volatility and jump-diffusion approach is to say that there is
risk-aversion and risk-premia are associated with un-hedgeable risks, or
in other words with the incompleteness of the market.

When using a BS-implied volatility surface, we do not need to worry
what causes the skew. If there are risk-premia, they will already be
in the option price. However, there is a qualifying statement to be made:
the P\&L of a $\De$-hedged long option position until expiry is
\begin{equation}
\int_0^T\left(\Th(t) + \frac{1}{2}\Ga_F(t) F^2(t)
\hat{\sigma}^2(t)-r(t)f(t)\right)dt
\label{eq:PL}
\end{equation}
where $\hat{\sigma}(t)$ is the actual realised volatility at time $t$,
and $\Ga_F$ is the futures-$\Ga$. All the integrands are stochastic, and
we thus do not expect the option to be priced at the expectation
of all these parameters. Instead, we expect that the option is priced
at the expectation -- under some risk-neutral or, if necessary, risk-adjusted
measure -- of the entire integral Eq.~(\ref{eq:PL}). This price can be expressed
by means of a Black-Scholes implied volatility. Thus, the option can be priced
at some implied $\sigma$ which can, in Eq.~(\ref{eq:PL}), replace the stochastic
$\hat\si(t)$. The Black-Scholes implied
volatility allows us to use the Black-Scholes option pricing theory, which
is an {\it effective theory}, where specific expectations (or integrals)
of some of the underlying stochastic variables are extracted from the
current prices of the traded assets. The effective dynamics which results
is based on some of the sources of uncertainty being `effectively'
integrated out of the full stochastic theory. It can be shown
that DLVFs are risk-adjusted expectations of future instantaneous
volatilities \cite{12}.
In this way, DLVF models are {\it effective theories of
volatility} in the same way as static forward interest rates define an
effective theory for interest rates. This is an important point which we
will return to later on.

Beyond the volatility risk in Eq.~(\ref{eq:PL}) of the P\&L of a hedged options
position, there is an equally important risk of spending an unusually
long amount of time in regions of high $\Ga$. However, we should expect
that no-one is compensated for this option-specific risk, simply
because this can be diversified away by selling a portfolio of vanilla
options providing a reasonably constant $\Ga$ profile across spot prices
and times. This portfolio would still have vega exposure, but no
option-specific risk \footnote{~In fact there is some option-specific risk,
because when the implied vols change for some options, the $\Ga$-profile
of the portfolio changes. This is also a vega risk, because
it is caused by the change of implied volatilities.}.
We thus formulate the following observations.
\begin{obs}
The implied Black-Scholes volatilities compensate the option seller for
volatility risk, but not for option-specific risk.
\end{obs}

If we use implied Black-Scholes volatilities to price exotic
options, we ignore that the exotic may have specific risk that can not
be diversified away by trading listed options. In particular that risk,
because it is a specific risk attached to an un-quoted asset, can not
be implied from the market. Furthermore, the same argument holds for any
model calibrated only on vanilla options.
This leads to the key criterion for successful exotic model valuation:
\begin{obs}
An option for which there is no known market price must either
be hedgeable with quoted securities,
or otherwise will incur instrument-specific risk premia.
\end{obs}
Obviously, this condition is satisfied by any sensible model whenever reality
stays within the constraints of the model-assumptions. The problems with the
DLVF models, as we argue here, is that reality is too far removed from the
fundamental DLVF assumption of deterministic (local) volatility parameters.
This is underlined by empirical research on market data \cite{11}.
A trivial aside of the above is
\begin{obs}
An exotic option
model has to either have a plausible way of incorporating option-specific
risk-premia or permit a proof that a replicating trading strategy exists.
\end{obs}
We will show below that the DLVF model applied to a world of non-deterministic
volatility parameters does neither if the only allowed hedging and
calibration instruments are European vanilla options.
It does not matter whether we use a low or high-parametric SLVF model:
In low-parametric SLVF models we may create option-specific risk
that can not be hedged by other vanilla options, simply because the
parameter-family of LVFs do not capture the full volatility exposure of all
conceivable exotics. In a high-parametric SLVF, we will normally not have
enough vanilla options to calibrate to, and worse: even if we have a
continuum of vanilla options for all strikes and expiries, we can not calibrate
a continuum of SLVF parameters to that.
It is therefore relevant to initially construct the space of all options
we are willing to consider as hedging instruments in our model, and then
decide which SLVF world this puts us in.
Obviously, in an SLVF(II)
world, there always is a hedging strategy for everything, but it would
have to be shown that we are in such a world.

The next step is thus to provide a mechanism which proves
the existence of a hedge against stock and volatility risk
in an LVF framework. We choose a trinomial world. More generally, we
simply try to answer the question of which of the SLVF worlds must be
the real one, if the real asset price process is SLVF, and if the set
of all exotics we wish to consider are all exotics that can be
defined in this discrete world. It therefore leads naturally to an
answer to the question whether all well-defined exotics can
be hedged in the continuum limit, and whether this continuum limit
is an SLVF(I) or SLVF(II) world if we only consider vanilla options in the
hedge. What is important to
remember, however, is that we will perform the model calibration
to the vanilla European options market in the DLVF sense, i.e.~we
do not calibrate the local volatilities' variance and covariance rates.
This is because we want to test the viability of the DLVF model in 
a world where the asset price process is SLVF, and where only vanilla
options, the underlying, and funding instruments are considered as hedging
instruments.

\section{Hedging an Exotic Option in an SLVF world using a DLVF model}

\subsection{Geometry of the tree}

A discrete local volatility tree with pre-determined state space needs
to be at least trinomial in order for the discrete process to
recombine \footnote{~Implied binomial trees are possible, but the state space
(i.e.~the up and down steps and the time steps) depends on synthetic options
which are very short-dated and forward starting a long time in the future.
We prefer not to import the complications of the market-calibration into the
construction of the state-space, so that volatility changes leave the
state-space unchanged. Also, the trinomial approach is much simpler, as the
Eqs.~(\ref{eq:trinomial-geometry}) leave exactly one free local parameter:
the local volatility.}.
Since we assume that interest rates can be hedged separately, we work in a
zero-interest-rate tree (the appropriate change of numeraire is thus
implicitly assumed), which we can specify with the following state-space
and transition probabilities
\bea
\nonumber
S_n^i=S_0e^{iu}\comma &-n\le i\le n\\
\nonumber
p\left(S_n^i\rightarrow S_{n+1}^{i-1}\right)&=\frac{\alpha_n^i}{1+e^{-u}}\\
\nonumber
p\left(S_n^i\rightarrow S_{n+1}^i\right)&=1-\alpha_n^i\\
p\left(S_n^i\rightarrow S_{n+1}^{i+1}\right)&=\frac{\alpha_n^i}{1+e^{u}}
\label{eq:trinomial-geometry}
\eea
where $S_0$ is the initial value of the stock price.
At any step, the stock price
change by a factor $e^{\pm u}$ or stay the same. $S_n^i$ is the stock
price at the $n^{\rm{th}}$ non-trivial time-slice after $i$ net up-moves.
The choice of notation
for the transition probabilities is an arbitrary parameterisation such that
the sum of the probabilities to leave a given node is one, and the 
stock price is a martingale, and the $\alpha$'s are strictly
monotonically increasing functions of the local volatilities.
The choice of the `global' tree-parameter $u$ puts an upper
bound on the possible local volatilities in this world, because, in order
to preserve positivity of the transition probabilities, we require
$0\le \alpha\le 1$. This does not put any fundamental restrictions on the
conclusions we will draw from analysing this discrete model \footnote{~From
$\hat{E}\left[\frac{S_{t+1}}{S_t}\right]=1$ and $\hat{E}\left[
\left(\frac{S_{t+1}}{S_t}-1\right)^2\right]=e^{\si^2\De t}-1$ we get
$\alpha_\tau^i=e^u\frac{e^{\left(\si_\tau^i\right)^2\De t}-1}{\left(e^u-1
\right)^2}$,
and the constraint $\alpha_\tau^i\le 1$ means that all local volatilities
have to satisfy
$\left( \si_\tau^i \right)^2\le\frac{1}{\De t}\ln(2\cosh u-1)$.}.
We make no assumptions about the process underlying the evolution of the
local volatility parameters $\al_n^i$. For example, restrictions on the
drift of local volatilities have to be posed by the condition that,
if local-volatility gadgets are attainable, the local volatilities and
the stock price have to be jointly martingale in order to avoid arbitrage
\cite{12}. However, we do not need to place any restrictions on the
evolution of the volatilities, except that they are bounded in the
interval $\left[0,u\right]$ \footnote{~A necessary condition for this is
that the volatilities of local volatilities are square-integrable over time
and that the modulus of the drift are integrable over time,
although these two conditions are not sufficient.}.

\subsection{Paths and vanilla options on the tree}

The set of all paths ${\mathscr{P}}$ has $3^n$ elements for a tree with $n$
time-slices. Define a path as the $n$-element sequence of the
space-slices the path visits
\[
\pi=\left(\pi_1,\pi_2,\ldots,\pi_n\right)
\comma
\left(\pi_0=0,
\left(\pi_i-\pi_{i-1}\right)\in \left\{-1,0,1\right\}\right)\,.
\]
We will also use the alternative definition, the $n$ element sequence of
steps (1=up, -1=down)
\[
\tilde{\pi}=\left\{\tilde{\pi}_1,\tilde{\pi}_2,\ldots,\tilde{\pi}_n\right\}
\comma\tilde{\pi}_i\in\left\{-1,0,1\right\}\,.
\]
The realised stock-path $\Pi(m)$ up to time $m$ gives a filtration on the
set of paths ${\mathscr{P}}$. We define
${\mathscr{P}}\left[\Pi(m)\right]\subset{\mathscr{P}}$ as the set of the paths
whose first $m$ steps match the realised path
$\Pi(m)$, and the final stock-value of the realised path as
$S_0e^{\Pi_m(m)u}$. We also define the set of directed paths
${\mathscr{D}}\left[m,i,\tau,j\right]=\left\{\pi\in{\mathscr{P}}\,\,,\,\,\,\pi_m
=i\,,\,\,\pi_\tau=j\right\}$, which means ${\mathscr{D}}\left[m,i,\tau,j\right]$
is the set of all paths which at time $m$ give a stock price of $S_0e^{iu}$
and at time $\tau$ a stock price of $S_0e^{ju}$.

Let us define the {\em future cone of a node} as the part of the tree which
can be reached from that particular node.
The question is whether it is possible
to hedge any path-dependent derivative when, at any time-step
$m\in\left\{0,1,\ldots,n-1\right\}$, the local volatilities
$\alpha_k^i$ in the future cone of $(m,\Pi(m))$ are not known with certainty,
and we only use European vanilla options as hedge instruments?

The restriction to use only European options is equivalent
to using only Arrow-Debreu options. On the lattice-world, this is trivially
true, whereas in the continuous world, some limiting procedure is necessary.
The path-probabilities,
as measured from time $m$, are
\bea
\nonumber
&p&\left(\pi,m,\Pi(m)\right)=
{\mathbf{I}}_{\pi\in{\mathscr{P}}[\Pi(m)]}\times\\*
\nonumber
&&\prod_{j=m}^{n-1}\left(
{\mathbf{I}}_{\tilde{\pi}_{j+1}=1}\frac{\alpha_j^{\pi_j}(m)}{1+e^{-u}}+
{\mathbf{I}}_{\tilde{\pi}_{j+1}=0}\left(1-\alpha_j^{\pi_j}(m)\right)+
{\mathbf{I}}_{\tilde{\pi}_{j+1}=-1}\frac{\alpha_j^{\pi_j}(m)}{1+e^{u}}\right)
\eea
where ${\mathbf{I}}_x$ is the indicator function which is
equal to 1 if $x$ is true and zero otherwise.

The Arrow Debreu price $A_\tau^i(m)$ is the price at time $m$ of the
derivative paying one unit if the stock-price at time $\tau$ is $S_0e^{iu}$,
i.e.~if the future realised path satisfies 
$\Pi(\tau)\in{\mathscr{P}}\left[\Pi(m)\right]\cap{\mathscr{D}}\left[0,0,\tau,j
\right]$.
Since $p\left(\pi,m,\Pi(m)\right)$ is zero whenever $\pi\notin
{\mathscr{P}}\left[\Pi(m)\right]$,
\[
A_\tau^i\left(m,\Pi(m)\right)=\sum_{\pi\in{\mathscr{D}}[0,0,\tau,j]}
p\left(\pi,m,\Pi(m)\right)\,,\,\,\,\tau> m.
\]
Note that the continuum limit of a discrete Arrow-Debreu price is an
ill-defined concept. Instead, a given continuous-world definition of
an Arrow-Debreu price needs to be matched to a discretisation-dependent
linear combination of discrete-world Arrow-Debreu prices. This will
then converge properly to the discrete-world Arrow-Debreu price. This is
normal procedure for path-integration problems, and is implicitly done
for any tree-model.

\subsection{Path-dependent options}

The most general path dependent option $f$ has a final pay-out of 
$X=X\left(\Pi(n)\right)$. In a similar fashion as for the Arrow-Debreu
options, we need to ensure that this definition provides a correct continuum
limit for the path integral. Thus, for the option to approximate a
continuum-world exotic,
$X=X\left(\Pi(n)\right)$ is a discretisation-dependent function.
The question is whether such a path-dependent
option can always be hedged with a strategy involving only the stock 
and Arrow-Debreu options, or whether such a hedge becomes valid at least
in the continuum limit. Under risk-neutral valuation, the value of
the path-dependent option at time $m$ is
\[
f\left(\Pi(m)\right)=\sum_{\pi\in{\mathscr{P}}}p\left(\pi,m,\Pi(m)\right)
X(\pi)\,.
\]
To determine the hedge portfolio, we have to know how many independent
securities there are available in this market. There are two
constraints on the Arrow Debreu prices
\[
\sum_{i=-\tau}^\tau A_\tau^i\left(m,\Pi(m)\right)=1\\
\]
\begin{equation}
\sum_{i=-\tau}^\tau A_\tau^i\left(m,\Pi(m)\right)e^{iu}=e^{\Pi_m(m)u}\,,
\label{eq:constraints}
\end{equation}
valid for all $\tau,m\,;\,\,\tau> m$. The first condition derives from
the fact that the stock-price at time $\tau$ will with certainty be on one
of the nodes of that time-slice, and the second simply says that a portfolio 
of Arrow Debreu prices which pays exactly the stock-price at any node at time
$\tau$ is equivalent to a portfolio holding only the stock. Henceforth, for
notational simplicity, we shall simply write $\Pi(m)$ instead of 
$\left(m,\Pi(m)\right)$, as the time can be inferred from the cardinality of
the representation of $\Pi(m)$. 
At time $m$, the most general portfolio $\Lambda$ short $f$ can thus be
represented as
\[
\Lambda(\Pi(m))=-f(\Pi(m))+\omega(\Pi(m))S(\Pi(m))+
\mbox{~~~~~~~~~~~~~~~~~~~~~~~~~~~~~~~~~~~~~~}
\]
\begin{equation}
\quad\sum_{t=m+1}^n\left(\sum_{i=\Pi_m(m)-(\tau-m)+2}^{\Pi_m(m)+(\tau-m)}
w_\tau^i(\Pi(m))A_\tau^i(\Pi(m))\right)\,.
\label{eq:portfolio-general}
\end{equation}
Note that the somewhat strange limits in the inner sum are there in order
to avoid specifying weights for worthless and redundant Arrow Debreu options.
Specifying
the entire time-indexed functionals $\omega(\Pi(m))$ and $w_\tau^i(\Pi(m))$
for all $m$ and
$\Pi(m)$ then defines a trading strategy. If there exists a hedged
trading strategy, it must be completely determined by the specification of
these functional. Note that we
assumed we are in a zero-interest-rate environment, so we have disregarded
the bond/deposit/loan in our portfolio, and the constraint that the
portfolio should be self-financing does not apply in the usual sense.
Instead, all we require from a hedged strategy is
\[
\Lambda(\Pi(m))=\Lambda'(\Pi(m+1))\,\,\,\forall\Pi(m+1)\in{\mathscr{P}}
\left[\Pi(m)\right]\,,
\]
where $\Lambda'$ is the portfolio one time-step ahead,
but before re-hedging. $\La'(\Pi(m+1))$ thus has the same portfolio weights as
$\Lambda(\Pi(m))$.
As time moves on one step, the
following relevant market parameters will have changed:
\begin{itemlist}
\item the stock price
\item the local volatilities in the future cone of $\Pi_m(m)$.
\end{itemlist}
Given that we assume that we know the volatility $\alpha_m^{\Pi_m(m)}$
with certainty (and given that, after $\De$ and $\Ga$-hedging in the
trinomial world we are in fact
not exposed to it at all), we can work out how many constraints and free 
variables there are to work out the hedge at every step.

\subsection{Constructing the hedge; tractor options}

Any path-dependent option (and therefore {\it any} option) can be written as
a linear combination of what we shall call tractor options. A tractor options
${\mathscr{T}}_\pi$ pays out one unit if and only if the stock-price up to time
$n$ has followed the path $\pi$. In the continuous world, we could call the
payoff of any contingent claim (exotic or otherwise) a functional on the space
of continuous functions $S(t)$. Each option valuation can be expressed as an
infinite-dimensional integral over the set of continuous functions on the
time to maturity $C\left([0,T]\right)$, with the corresponding pay-off
functional as the integral kernel, which leads to the path-integral
representation of risk-neutral option valuation. It is possible that the
analysis which follows can be consistently formulated in this continuous
framework, but the path-integral can also be represented as a limiting 
case of path-integrals over functions on discrete image spaces as long as
the mapping from the continuous-world derivative to the corresponding
discrete-world linear combination of tractor options is done in a way that
demonstrably converges to the continuous-world pay-off.

We are going this route of replacing the path-integral by a sum of paths in a
discrete world, and then taking the continuum limit.
The question of whether all exotic options can be hedged can now
be re-phrased as the question of whether all tractor options can be
hedged. Note that the value of a tractor option at time $m$ is simply
\[
{\mathscr{T}}_\pi\left(\Pi(m)\right)=p\left(\pi,\Pi(m)\right)\,,
\]
so we see that we have in fact been using tractor options already in the
discussion so far, and that -- in the language of path-integration -- the
tractor option prices form the probability measure on the function 
space for the functions $S(t)$. So we see that ${\mathscr{P}}$ is the set of
all states of the world, and the set of all tractors generates the largest
possible $\si$-algebra on that space. ${\mathscr{T}}:{\mathscr{P}}\rightarrow
[0,1]$ is our probability measure on that space, and
$\left( {\mathscr{P}},\si \left( \left\{{\mathscr{T}}_{\pi_1}\right\},
\left\{{\mathscr{T}}_{\pi_2}\right\},\ldots,
\left\{{\mathscr{T}}_{\pi_{3^n}}\right\}\right),{\mathscr{T}}\right)$
is the relevant probability space for valuing options in a
risk-neutral valuation framework. It can also be seen that our above
definition of an Arrow-Debreu price is itself a linear combination of 
tractor prices
\[
A_\tau^i(\Pi(m))=\sum_{\pi\in{\mathscr{D}}[0,0,\tau,j]}{\mathscr{T}}_\pi(\Pi(m))\,,
\,\,\tau>m\,,
\]
which means that the Arrow Debreu prices are the probabilities associated
with a particular partition of the state space. Note that this equation
holds independently of the local volatilities. We may ask whether the
reverse is true, i.e.~whether we can write the tractor as a linear
combination of Arrow-Debreu options. The answer is strictly no, as can be
easily seen by the fact that there are many more tractors than Arrow-Debreu
prices, and there are probably no more constraints on the tractor prices
than we had for the Arrow-Debreu prices in Eq.~(\ref{eq:constraints}).
So there is no 
general static hedging strategy possible for all tractors. In set-theoretical
notation, the $\si$-algebra generated by the sets ${\mathscr{D}}[0,0,\tau,j]
\cap{\mathscr{P}}\left[\Pi(m)\right]$ (relevant to observing which Arrow-Debreu
option pays off) is much smaller than the $\si$-algebra generated by the
sets  $\{\pi_i\}\,,\,\,\pi_i\in{\mathscr{P}}\left[\Pi(m)\right]$ (relevant
to observing which tractor option pays off). Let us again consider the
portfolio Eq.~(\ref{eq:portfolio-general}). How many free parameters are there
to adjust the hedge with? It is easy to convince oneself that there will
be $(n-m)^2$ independent Arrow-Debreu weights \footnote{~Since
$\sum_{\tau=m+1}^n\left(\sum_{i=-(\tau-m)+2}^{\tau-m}1\right)=(n-m)^2$.}
and one stock-weight. The number of constraints depends on our hedging
ambitions. We could attempt to hedge against the volatilities to any
order, which we will demonstrate to be impossible for any tree of more
than two steps. However, this is only necessary in a jump-diffusion world.
If we want to be hedged against volatility to second order, as we should
be in an SLVF world, the constraints are
\bea
\nonumber
&\Lambda'\left(\Pi(m)\otimes\{1\}\right)=
\Lambda'\left(\Pi(m)\otimes\{0\}\right)=
\Lambda'\left(\Pi(m)\otimes\{-1\}\right)\\[2mm]
\nonumber
&\dfrac{\partial\Lambda'\left(\Pi(m)\otimes\{1\}\right)}{\partial\alpha_i^j}=
\dfrac{\partial\Lambda'\left(\Pi(m)
\otimes\{-1\}\right)}{\partial\alpha_i^j}\\[2mm]
\nonumber
&\dfrac{\partial\Lambda'\left(\Pi(m)\otimes\{0\}\right)}{\partial\alpha_i^j}=0
\\[2mm]
\nonumber
&\dfrac{\partial^2\Lambda'\left(\Pi(m)\otimes\{0\}\right)}
{\partial\alpha_i^j\partial\alpha_k^l}=0\\[2mm]
\nonumber
&\forall i=m+1,\ldots,n-1\,,\,\,j=\Pi_m(m)-(i-m),\ldots,\Pi_m(m)+(i-m)\\[2mm]
&\forall k=m+1,\ldots,n-1\,,\,\,j=\Pi_m(m)-(k-m),\ldots,\Pi_m(m)+(k-m)
\label{eq:K2}
\eea
The first equation enforces $\De$ and $\Ga$ hedging, the second equation
enforces cross-terms between volatility and stock, the third equation
enforces `local-vol-parameter' vega hedging, and the fourth
enforces all the second derivatives with respect to the volatilities'
parameterisation. Note that in the second line in Eq.~(\ref{eq:K2}) we do
not require the first derivatives to be zero, as we do in the third
line, as this would also enforce a hedge against 
the option price derivative $\frac{\partial^3f}{\partial S^2\partial 
\al_i^j}$, which is more than we need. Because
the derivative $\frac{\partial}{\partial S}$ has no unique discrete
representation on the trinomial tree, the derivative $\frac{\partial^2f}
{\partial S\partial \al_i^j}$ also has no unique representation. However,
we need not dwell too much on this technicality, as we will see that even our
potentially too undemanding set of Eqs.~(\ref{eq:K2}) can not be satisfied in
a three-step world and beyond.

If we want to be hedged against local volatility movements only to
first order, the constraints are
\bea
&\Lambda'\left(\Pi(m)\otimes\{1\}\right)=
\nonumber
\Lambda'\left(\Pi(m)\otimes\{0\}\right)=
\Lambda'\left(\Pi(m)\otimes\{-1\}\right)\\[2mm]
\nonumber
&\dfrac{\partial\Lambda'\left(\Pi(m)\otimes\{0\}\right)}{\partial\alpha_i^j}=0
\\[2mm]
&\forall i=m+1,\ldots,n-1\,,\,\,j=\Pi_m(m)-(i-m),\ldots,\Pi_m(m)+(i-m)
\label{eq:K3}
\eea
We will show that the constraints Eqs.~(\ref{eq:K3}) can always be satisfied.
The algebra of the problem is 
very involved, and from the four-step or even three-step world upwards the
equations are far too many to analyse all possible tractor hedges without the
help of an algebraic calculations program.
An example Mathematica$^{\rm{TM}}$ notebook can be requested from the author.

\newpage
\subsection{Example: A two-step world}

As a first example, we consider a two-step tree where we are short $K$
tractor options $f_{\{1,-1\}}$. Later, we shall also consider the most
general exotic option in this two-step trinomial world. The zero-time 
hedged portfolio is
\[
\Lambda(\emptyset)=
-f(\emptyset) +\omega(\emptyset)S_0
+w_1^1(\emptyset) A_1^1(\emptyset)\mbox{~~~~~~~~~~~~~~~~~~~~~~~~~~~~~~}
\]
\[
+w_2^2(\emptyset) A_2^2(\emptyset)
+w_2^1(\emptyset) A_2^1(\emptyset)
+w_2^0(\emptyset) A_2^0(\emptyset)
\]
After one step, before re-hedging, we have
\[
\Lambda(\Pi(1))=
-f(\Pi(1))
+\omega(\emptyset) S_1^{\Pi_1(1)}
+w_1^1(\emptyset) A_1^1(\Pi(1))\mbox{~~~~~~~~~~~~~~~~~~~~}
\]
\[
\quad\mbox{~~~~~~~~~}
+w_2^2(\emptyset) A_2^2(\Pi(1))
+w_2^1(\emptyset) A_2^1(\Pi(1))
+w_2^0(\emptyset) A_2^0(\Pi(1))
\]
In full
\bea
\nonumber
\Lambda(\{1\})&=&-K\frac{\alpha_1^1(1)}{1+e^{-u}}+\omega(\emptyset)S_0e^u+
w_1^1(\emptyset)\\
\nonumber
&&\quad +w_2^2(\emptyset)\frac{\alpha_1^1(1)}{1+e^u}
+w_2^1(\emptyset)(1-\alpha_1^1(1))
+w_2^0(\emptyset)\frac{\alpha_1^1(1)}{1+e^{-u}}\\
\nonumber
\Lambda(\{0\})&=&\omega(\emptyset)S_0
+w_2^1(\emptyset)\frac{\alpha_1^0(1)}{1+e^u}
+w_2^0(\emptyset)(1-\alpha_1^0(1))\\
\nonumber
\Lambda(\{-1\})&=&\omega(\emptyset)S_0e^{-u}
+w_2^0(\emptyset)\frac{\alpha_1^{-1}(1)}{1+e^u}
\eea
If we attempt to hedge to second order, we solve in all the
Eqs.~(\ref{eq:K2}). Each of the portfolios depends only on one of the
future local volatilities, so to first order all the first
derivatives with respect to the local volatilities have to be zero,
which leads to
\[
w_2^1(\emptyset)=w_2^0(\emptyset)=0\comma w_2^2(\emptyset)=Ke^u
\]
After that, the condition
$\Lambda'(\{0\})=
\Lambda'(\{-1\})$
leads to $\omega(\emptyset)=0$, and 
$\Lambda'(\{0\})=
\Lambda'(\{1\})$
to 
$w_1^1=0$. So the only hedge we put on initially is $e^uK$ units of $A_2^2$.
We also see that this makes the portfolio worthless initially. If the first
step of the stock-price path is flat or down, the portfolio is patently
worth-less. If it is up, the portfolio has also value zero, regardless of
the actual value of the remaining local volatility in the future cone, 
$\alpha_1^1$. The hedge portfolio for this step is then trivially
calculated by
$\Lambda'(\{1,1\})=
\Lambda'(\{1,0\})=
\Lambda'(\{1,-1\})$. The solution is $\omega(\{1\})=-\frac{K}{S_0(e^u-1)}
\,\,,\,\,\,w_2^2(\{1\})=e^uK$.
So the entire hedging strategy was found easily. Note that the hedge
constraints were all linear in the local volatilities, because the
constraints are formulated at the last time-slice. Any exotic
option will have collapsed to an Arrow Debreu option by that time. Thus,
the $\si$-algebra generated by the sets ${\mathscr{D}}\left[0,0,1,j\right]
\cap{\mathscr{P}}\left[\Pi(m)\right]$ and the $\si$-algebra generated by
the paths still relevant after the first time-step (i.e.~when the hedging
conditions need to be matched) are identical. Thus we expect the two-step
world to provide a hedging strategy for all tractors. To demonstrate this,
we can work out the whole hedged trading strategy.
Let us write the general two-step exotic as a linear combination
\[
f=\sum_{i,j=-1}^1 K_{\{i,j\}}{\mathscr{T}}_{\{i,j\}}.
\]
Then the part of the hedging strategy relevant to the first time-step is
\bea
\nonumber
\omega(\emptyset)&=&\dfrac{
K_{\{0,0\}}-
K_{\{-1,1\}}+
e^u K_{\{-1,0\}}}
{(1-e^{-u})S_0}+
\dfrac{e^{2u}
K_{\{-1,-1\}}}{(1-e^u)S_0}\\[2mm]
\nonumber
w_1^1(\emptyset)&=&
K_{\{1,0\}}-
K_{\{0,1\}}-
e^u K_{\{0,-1\}}+
e^u K_{\{-1,0\}}\\[2mm]
\nonumber
w_2^0(\emptyset)&=&
K_{\{-1,1\}}-
(1+e^u)K_{\{-1,0\}}+
e^u K_{\{-1,-1\}}\\[2mm]
\nonumber
w_2^1(\emptyset)&=&
K_{\{0,1\}}-
(1+e^u)K_{\{0,0\}}+
e^u K_{\{0,-1\}}+(1+e^u)w_2^0\\[2mm]
\nonumber
w_2^2(\emptyset)&=&
K_{\{1,1\}}-
(1+e^u)K_{\{1,0\}}+
e^u K_{\{1,-1\}}+(1+e^u)w_2^1-e^u w_2^0
\eea
It is easy to check that with these parameters, the value of the portfolio
is indeed independent of all the future volatilities. In all the cases
the final hedge is trivial, and we will not demonstrate how to compute it.
The reason that the hedge is trivial on the last time-step is that all
the tractor options will have collapsed to Arrow-Debreu options.
For this reason the problem of finding 
an exactly replicating portfolio in the two-step world is solvable.
Note that a three-step world, however, collapses to a two-step world
after the first step, and one third of all the tractor options
become worthless at that time.
The remaining ones are still two-step tractors.
We will demonstrate by the example of a particular tractor option
that the general three-step exotic can not be replicated to second order
in  the stock-price and local volatilities with a dynamic
trading strategy of vanilla products.

\subsection{Hedging Three-Step Tractors}
\label{sec:threestep}

Imagine we want to hedge $K$ times the tractor ${\mathscr{T}}_{\{0,0,0\}}$
in a three-step world, i.e.~the exotic which pays off $K$ units only if
the stock never moves at all. For convenience, we will generally omit the 
arguments of $w$, and $\alpha$. Then the portfolios after one step are
\bea
\nonumber
\Lambda'(\{1\})&=&
w_1^1+\dfrac{\al_1^1}{1+e^{-u}}w_2^0+(1-\al_1^1)w_2^1+
\dfrac{\al_1^1}{1+e^u}w_2^2+\dfrac{\al_1^1\al_2^0}{(1+e^{-u})^2}w_3^{-1}\\
\nonumber
&&\quad +\dfrac{\al_1^1(1-\al_2^0)+(1-\al_1^1)\al_2^1}{1+e^{-u}}w_3^0\\
\nonumber
&&\quad +\left(\dfrac{\al_1^1(\al_2^2+\al_2^0)}{(1+e^u)(1+e^{-u})}+
(1-\al_1^1)(1-\al_2^1)\right)w_3^1\\
\nonumber
&&\quad +\dfrac{\al_1^1(1-\al_2^2)+(1-\al_1^1)\al_2^1}{1+e^u}w_3^2+
\dfrac{\al_1^1\al_2^2}{(1+e^u)^2}w_3^3+e^u\omega S_0\\[4mm]
\nonumber
\La'(\{0\})&=&(1-\al_1^0)w_2^0+\dfrac{\al_1^0}{1+e^u}w_2^1+
\dfrac{\al_1^0(1-\al_2^{-1})+(1-\al_1^0)\al_2^0}{1+e^{-u}}w_3^{-1}\\
\nonumber
&&\quad +\left(\dfrac{\al_1^0(\al_2^1+\al_2^{-1})}{(1+e^u)(1+e^{-u})}+
(1-\al_1^0)(1-\al_2^0)\right)w_3^0\\
\nonumber
&&\quad +\dfrac{\al_1^0(1-\al_2^1)+(1-\al_1^0)\al_2^0}{1+e^u}w_3^1+
\dfrac{\al_1^0\al_2^1}{(1+e^u)^2}w_3^2\\
\nonumber
&&\quad +\omega S_0 -K(1-\al_1^0)(1-\al_2^0)\\[4mm]
\nonumber
\La'(\{-1\})&=&\dfrac{\al_1^{-1}}{1+e^u}w_2^0+\left(\dfrac{\al_1^{-1}
(\al_2^0+\al_2^{-2})}{(1+e^u)(1+e^{-u})}+(1-\al_1^{-1})(1-\al_2^{-1})\right)
w_3^{-1}\\
\nonumber
&&\quad +\dfrac{\al_1^{-1}(1-\al_2^0)+(1-\al_1^{-1})\al_2^{-1}}{1+e^u}w_3^0+
\dfrac{\al_1^{-1}\al_2^0}{(1+e^u)^2}w_3^1+e^{-u}\omega S_0
\eea
 
Let us try to find hedge ratios so as to satisfy the Eqs.~(\ref{eq:K2}).
We will not illustrate all the tedious algebra involved, but we present
one possible way to proceed, giving only some intermediate steps in
the solution. First consider
$\frac{\partial\La'(\{0\})}{\partial\al_2^1}=0$,
which leads to
$w_3^2=-e^uw_3^0+(1+e^u)w_3^1$.
Next,
$\frac{\partial\La'(\{0\})}{\partial\al_2^0}=0$,
gives
$w_3^1=-(1+e^u)K-e^uw_3^{-1}+(1+e^u)w_3^0$.
$\frac{\partial\La'(\{0\})}{\partial\al_2^{-1}}=0$
leads to
$w_3^0=(1+e^u)w_3^{-1}$.
Next,
$\frac{\partial\La'(\{1\})}{\partial\al_2^{-2}}=
\frac{\partial\La'(\{-1\})}{\partial\al_2^{-2}}$
yields
$w_3^{-1}=0$.
Demanding that 
$\frac{\partial\La'(\{1\})}{\partial\al_2^2}=
\frac{\partial\La'(\{-1\})}{\partial\al_2^2}$
then gives
$w_3^3=-(1+e^u+e^{2u})(1+e^u)K$.
The
simplified solutions so far are thus
$w_3^0=w_3^{-1}=0$, $w_3^1=-(1+e^u)K$, $w_3^2=-(1+e^u)^2K$, and
$w_3^3=-(1+e^u+e^{2u})(1+e^u)$.
This settles the hedge ratios for the longest-living Arrow-Debreu
prices. Next, we satisfy
$\frac{\partial\La'(\{1\})}{\partial\al_1^1}=
\frac{\partial\La'(\{-1\})}{\partial\al_1^1}$
by setting
$w_2^2=-e^uw_2^0+(1+e^u)w_2^1-e^u\al_2^0K$.
The fact that the hedge ratio is dependent on a local volatility is
not really a problem. We need to be careful, however, not to treat
any $\al$'s which appear in a hedge ratio as variable
when calculating other hedge-ratios
because the implicit assumption is always that, while the stock
price or the volatility are allowed to change instantaneously, the
hedge ratios must stay constant until after the information on
the new levels of stock price and volatilities has become known.
With this in mind, the condition
$\frac{\partial\La'(\{1\})}{\partial\al_1^{-1}}=
\frac{\partial\La'(\{-1\})}{\partial\al_1^{-1}}$
leads to
$w_2^0=K\al_2^0$.
Now, without demanding to fulfil the remaining
Eqs.~(\ref{eq:K2}), we are already lead to an inevitable violation of
one part of Eqs.~(\ref{eq:K2}), namely that
$\frac{\partial\La'(\{1\})}{\partial\al_2^{0}}-
\frac{\partial\La'(\{-1\})}{\partial\al_2^{0}}=\frac{K}{1+e^u}(\al_1^{-1}
-e^u\al_1^1)$.
There is no further hedge ratio that we can adjust to
remedy this problem \footnote{~~In fact, we can just stubbornly proceed
to solve all the other conditions in Eqs.~(\ref{eq:K2}), namely
$\frac{\partial\Lambda'(\{1\})}{\partial\al_1^0}=0$
and
$\La'(\{1\})=
\La'(\{0\})=
\La'(\{-1\})$,
which give the hedge ratios
$w_2^1=(1+e^u)\al_2^0K$,
$w_2^2=(1+e^u)^2\al_2^0K$,
$w_1^1=0$, and $\omega=\frac{1-\al_2^0}{1-e^{-u}}K$, but this will not
change the nature of the problem.}.

Now we can of course do the same analysis for all the 27 tractor options
in this small trinomial world, but we do this with Mathematica$^{\rm{TM}}$.
It turns out that in general, all but one of the second-order
conditions of the Eqs.~(\ref{eq:K2}) can be satisfied
and hedging to first order is possible.

\subsection{Analysis of Larger Trinomial Worlds}

When analysing larger trinomial worlds, we quickly run into constraints in
terms of processing power.
Our Mathematica$^{\rm{TM}}$ implementation frequently runs out of memory
and needs
copious manual intervention when analysing a five-step trinomial world.
Nevertheless, some valuable observations can be made even with our calculations
limited to a five-step and smaller trinomial worlds.

The number of violated constraints in Eqs.~(\ref{eq:K2}) increases very
quickly in going from a three-step to a four-step and five-step world.
Obviously, the constraints arising in an $n$-step world are a subset 
of the constraints in any larger trinomial worlds, so in general we can not
hope to find an exact second order hedge in any larger trinomial world.
The question is, however, whether these terms matter, or whether their
number becomes less and less significant when compared to the relevant
terms.

Let us define the number of un-satisfiable constraints in Eqs.~(\ref{eq:K2})
for a particular tractor ${\mathscr{T}}_{\pi}$ as $\nu_\pi$. Then we can 
for example measure the number
\[
N_n=\sum_{\pi\in{\mathscr{P}}_n}\nu_\pi,
\]
where we have given the set ${\mathscr{P}}$ a subscript $n$ to indicate
how many steps the trinomial world has. We find
\[
\{N_1,N_2,\ldots\}=\{0,0,21,210,1344,\ldots\}
\]
Note that on a 300 MHz/192 MB personal computer, the last number took about
12 hours to compute in Mathematica$^{\rm{TM}}$. The estimated time for
the next element in the series is 160 hours, needing constant human
intervention. So we have to make our conclusion from this limited sequence.
In fact this sequence already indicates what kind of severity of mishedge we
might expect in the continuum limit:
In an $n$-step tree, there are $n^2-1$ local
volatilities in the future cone of the origin, which means there are $n^2$
stochastic parameters to consider. This gives $\frac{n^2(n^2+1)}{2}$ P\&L terms
in the continuous trading equation arising from the quadratic variations of
the stochastic variables alone.
Assuming that the total vega-risk of vanilla options should not be affected by
the discretisation, the relative importance of
each particular quadratic term that can not be hedged is thus weighed down
by a factor $1/(n^2(n^2+1))$ on average, when we compare different trinomial
worlds. There are $3^n$ tractors in each $n$-step trinomial world, so the
relative importance of each tractor, when defining an exotic option, will
be weighed down by a factor $3^{-n}$.
As a rule of thumb, the relevance of all unsatisfied hedge constraints
stays constant if, for large $i$, $N_i\propto 3^ii^2(i^2+1)$. Starting
with $N_3=21$, the corresponding estimate of $N_4$ would be
$3N_3\frac{4^2(4^2+1)}{3^2(3^2+1)}=190.4$, which is somewhat below the
actual $N_4$, while the estimate of $N_5$ would be
$9N_3\frac{5^2(5^2+1)}{3^2(3^2+1)}=1362$, which slightly larger than
the actual $N_5$. With the limited sequence we were able to compute, it
thus seems very plausible that the relative importance of the
un-hedgeable terms does not disappear in the continuum limit.

Lastly, we note that it is of course possible to construct `exotics',
i.e.~linear combination of tractors, which do not exhibit any quadratic
or linear volatility exposure after an optimal hedge is put in place.
A trivial example are the European vanilla
options themselves; another, less trivial, example are the tractors
${\mathscr{T}}_{\{1,1,1,\ldots,1,1,i\}}\,\,i\in\{-1,0,1\}$ and
${\mathscr{T}}_{\{-1,-1,-1,\ldots,-1,-1,i\}}\,\,i\in\{-1,0,1\}$, and linear
combinations thereof. Also, arbitrary tractors may often have more complicated
linear combinations which are insensitive to local volatilities to second
order. In the set of all exotic options, however, these will form a set
of measure zero: The linear vector space spanned by the set of all tractor
options is the space of all contingent claims on the underlying asset.
The space spanned by all independent linear combinations which have no
volatility exposure (to second order) after hedging with vanilla options
is a subspace of that set. If
any tractor option at all has un-hedgeable local-vega risk, this
subspace is a true subspace, and thus forms a set of measure zero in the
space of all contingent claims. Thus we have shown that -- out of all
conceivable contingent claims -- almost all are unhedgeable with a trading
strategy using vanilla options only. The severity of the problem is not
mitigated in the continuum limit.

\section{Discussion and Conclusions}

We have demonstrated that, in an arbitrage-free market, pricing exotic
options with a DLVF model (implied
tree) is inconsistent with the assumption of unknown future volatilities,
inasmuch as the LVF implied in the European vanilla options market does not
fully compensate for irreplicable volatility
risks exhibited by exotics if vanilla options are the only instruments
used for hedging volatility exposure.
Intuitively, we can interpret this result in the following manner:
Local volatility gadgets exist, but they do
not rid us of second-order local volatility
exposure, which generates steady additional (positive or negative) P\&L over
time. This P\&L is not priced in \footnote{~~Furthermore, the implied tree
generally gives no description of how the best possible hedge can be
attained even within these limitations. This, however, is a purely practical
issue.}.

By contrast, a market where the volatility process has only linear variation
does not give us the second-order P\&L terms, but is not arbitrage-free.
A market where the volatility has no variation at all gives no vega-P\&L
problems at all, and is arbitrage free. It is under the assumption
of this market that the DLVF model is constructed, which is theoretically
self-consistent, but not consistent with reality \cite{10,11}.

We stress that the lack of a second-order vega-hedge is not model-specific:
even a stochastic volatility model would create that, as long as the
volatilities which we try to imply are not substantially fewer than the
number of vanilla option prices one calibrates the model to. However, in
stochastic volatility models and jump-diffusion models the onus of pricing
in volatility risk is not on the implied volatilities alone. Instead, the
price of such risks can be calibrated to the market and particular
model parameters exist to absorb this calibration explicitly. Low-parametric
SLVF models do therefore not require the volatility risk to be {\it effectively}
integrated out and -- in some mysterious way -- absorbed into the DLVF
calibration.

We have only considered LVF models. It should be noted, however, that what
we pointed out here is a general concern which needs to be addressed for
any asset-price model one is willing to consider: The model must price
exotics consistently with the market parameters that one is calibrating
the model to. As we have seen here, it is not always easy to convince
oneself whether or not this demand is actually met by a proposed model.
All we have discovered in this paper is that it is not possible to
satisfy this condition when using an LVF model calibrated to vanilla option
prices if (a) we only use the asset and vanilla options as hedge
instruments, (b) volatilities are not completely deterministic,
(c) the market is arbitrage-free, and (d) the number of local volatilities
is of the order of the number of vanilla options available for model
calibration. As this is a central conclusion, we sum it up as another
\begin{thm}
Vanilla options do not complete the market in the most general SLVF model.
\end{thm}

However, it is possible that no single model price process satisfies this
demand if we use one and the same calibration recipe regardless of what
contingent claim we want to price. Thus, we must
find a proof for a different model that it does price
exotics in line with the vanilla options market. Possible resolutions of
this dilemma might be
\begin{itemlist}
\item
A low-parametric LVF surface calibration which includes volatility risk in
what must essentially be an SLVF model.
\item
A different calibration using quoted exotic options as well as vanilla
European options. In this case we must also allow the use of these
exotics as hedge instruments.
\item
A jump-diffusion model, or a combination of jump-diffusion with SLVF
or with DLVF as in ref.~\cite{16}
\item
A model which makes no self-consistent assumptions about the underlying
process at all, but proves in other ways that it prices (maybe only
specific) exotics `in line with' the market. This amounts to finding
a model-independent hedging strategy \footnote{~~Since
one makes no assumptions about
the underlying process, this model can not be general enough to allow
us to price all conceivable exotics. As an example of the concept just note
that it is trivial to price European options with arbitrary payoffs from
vanilla European options in a model-independent way.}.
\end{itemlist}
There is certainly no guarantee that any of the first three resolutions
would work for all possible exotics, and the exit route proposed in the last
item is likely to involve substantially different derivations for each
different class of exotic options. Also, there is no guarantee that a
model-independent hedging-solution exists for any particular exotic option.

Faced with the task of pricing a particular exotic option while not having
a bullet-proof hedging model, it must
be of paramount concern to know the un-priced risks, whatever the model.
The DLVF model gives us no information whatsoever on this.
However, ideas such as the `uncertain volatility
models' of Avellaneda, Levy, and Par\'as, and of Lyons
have emerged in an attempt to avoid the perils of trusting a single
underlying-process to price all exotics correctly \cite{17,17a,18}.
The main premise of
their approach is that the exotic pricing should be performed as
model-independently as possible, by using optimum amounts of standard traded
options to statically hedge out as much of the payoff as possible;
thereafter the residual is delta-hedged in order to obtain the tightest
possible worst/best price spread for the option within the assumption of
a `certainty interval' or confidence interval
for future volatilities \cite{17,17a,18,19}.

In summary, we should never trust a model-process of the underlying to price a
particular exotic correctly, unless we can prove this to be the
case \footnote{~~The best way to prove that the model works is by analysing
the hedging strategy that the model would suggest. While we saw in this paper
that it is difficult to extract a hedging strategy from an DLVF model,
other models may lend themselves more naturally to an interpretation in
terms of the proposed hedging method.}.
The way in which the DLVF model
is often trusted to even price {\em all conceivable} exotic options
correctly -- while we have shown that the set of correctly priced options
is infinitely smaller (of measure zero in the space of all exotics) --
should serve a warning.
The effective theory works for almost none of the conceivable derivatives.

As a by-product of the proof, we have also found that any tradable
asset, if it is stochastic, must have at least quadratic variation
-- and will therefore produce a $\Ga$-like P\&L on instruments with
non-linear exposure to that asset -- if the market is arbitrage-free.
This result complements the observations of the HJM model and of
Derman and Kani that the quadratic variation of their particular
diffusive models determines the drift completely if the condition of no
arbitrage is imposed \cite{14,12}. Even in the context of jump-diffusion models
as for example in \cite{16}, the drift is required to be deterministic
for the same reason. Since the drift cannot be stochastic in the absence
of arbitrage, we must either have at least quadratic variation or full
determinism.

\end{document}